\documentclass[aps,pra,reprint,nofootinbib,twocolumn,superscriptaddress,showpacs,showkeys,longbibliography]{revtex4-1}

\usepackage{eurosym}
\usepackage{amsmath,amssymb,amstext}
\usepackage{ulem}
\usepackage[usenames,dvipsnames]{color}
\usepackage{graphicx}
\usepackage{braket}
\usepackage{natbib}
\usepackage{comment}
\usepackage{dcolumn}
\usepackage[english]{babel}
\usepackage{wasysym}
\usepackage{bm}
\usepackage[colorlinks,bookmarks=false,citecolor=blue,linkcolor=red,urlcolor=blue]{hyperref}

\usepackage{appendix}
\usepackage{CJK}

\begin{document}
\title{Determination of enantiomeric excess with chirality-dependent AC Stark effects in cyclic three-level models}
\author{Chong Ye}
\affiliation{Beijing Computational Science Research Center, Beijing 100193, China}
\author{Quansheng Zhang}
\affiliation{Beijing Computational Science Research Center, Beijing 100193, China}
\author{Yu-Yuan Chen}
\affiliation{Beijing Computational Science Research Center, Beijing 100193, China}
\author{Yong Li}\email{liyong@csrc.ac.cn}
\affiliation{Beijing Computational Science Research Center, Beijing 100193, China}
\affiliation{Synergetic Innovation Center for Quantum Effects and Applications, Hunan Normal University, Changsha 410081, China}

\date{\today}

\begin{abstract}
  Determination of enantiomeric excess is important and remains challenges.
  We theoretically propose a new spectroscopic method for this issue based on the chirality-dependent AC Stark effects in cyclic three-level models under the three-photon resonance condition. The enantiomeric excess of the chiral mixture is determined by comparing the amplitudes of the two chosen AC Stark peaks in the Fourier transform spectrum of the induced polarizations, which are (approximately) proportional to the molecule numbers of the two enantiomers, respectively. Comparing with current spectroscopic methods based on the interference between the electric- and (usually weak) magnetic-dipole transition moments and/or with the need for enantio-pure samples, our method only involves electric-dipole transitions and does not require the enantio-pure samples. Therefore, it will give strong chiral signals and can be applied to the determinations of enantiomeric excess for chiral molecules whose enantio-pure samples are still challenging to achieve.

\end{abstract}

\maketitle

\section{Introduction}
Chirality is fundamentally important for the chirality-dependency of
many chemical and biological processes. Despite this,
determination of enantiomeric excess remains tremendous
challenges~\cite{PC1,PC2,PC3,PC4,PC5,Book1}. Traditionally, there are
some spectroscopic methods~\cite{CA1,CA2,CA3} for determination of enantiomeric excess, such as
circular dichroism, vibrational circular dichroism, and Raman optical activity.
These methods~\cite{CA1,CA2,CA3} are based on the interference between the electric-dipole transition
moments and the usually weak magnetic-dipole transition moments. Therefore, the related
chiral signals are usually weak.

Recently, based on the cyclic three-level model~\cite{SCP,PRL.87.183002,PRL.90.033001,PRA.77.015403,PRL.99.130403,JCP.132.194315,PRA.98.063401} of chiral molecules, enantiomer-specific microwave spectroscopy (EMS) method~\cite{Hirota,Nature.497.475,PRL.111.023008,PCCP.16.11114,ACI,JCP.142.214201,JPCL.6.196,JPCL.7.341,Angew.Chem.10.1002,KK}
has achieved great success in the determination of enantiomeric excess.
Two driving electromagnetic fields, resonantly coupling two electric-dipole transitions, give rise to an induced polarization corresponding to the third electric-dipole transition due to the mechanism of three-wave mixing in the framework of the cyclic three-level model. Since the product of the three electric-dipole transition moments
changes sign for different enantiomers, there is a phase shift of $\pi$ for the induced polarizations of the two enantiomers. In the chiral mixture to be detected, the induced polarizations from the two enantiomers will destructively interfere and thus the amplitude of the total induced polarization will reflect the molecule number difference between the two enantiomers. With the help of an enantio-pure sample with the same molecule number, one can determine the enantiomeric excess of the chiral mixture to be detected. The EMS method~\cite{Hirota,Nature.497.475,PRL.111.023008,PCCP.16.11114,
ACI,JCP.142.214201,JPCL.6.196,JPCL.7.341,Angew.Chem.10.1002,KK}
involves only the electric-dipole transition moments and thus is
considered as a sensitive, quantitative method for determination of enantiomeric excess~\cite{Hirota,Nature.497.475,PRL.111.023008,PCCP.16.11114,
ACI,JCP.142.214201,JPCL.6.196,JPCL.7.341,Angew.Chem.10.1002,KK}.

Similar to the traditional methods~\cite{CA1,CA2,CA3}, the enantio-pure samples are needed in the EMS method~\cite{Hirota,Nature.497.475,PRL.111.023008,PCCP.16.11114,
ACI,JCP.142.214201,JPCL.6.196,JPCL.7.341,Angew.Chem.10.1002,KK}. For some chiral molecular
species~\cite{Nature.497.475,ACI,PRL.111.023008,PCCP.16.11114},
the enantio-pure samples are commercial available.
However, it is still challenging to achieve the enantio-pure samples for most chiral molecular
species~\cite{1808.08642,PRL.121.173002,PRL.87.183002,PRL.90.033001,
PRA.77.015403,PRL.118.123002,PRL.99.130403,JCP.132.194315,PRL.84.1669,
PRA.65.015401,JCP.115.5349,JCP.119.5105,PRL.122.173202}.
In the two most widely used techniques for achieving the enantio-pure samples, chromatography and capillary electrophoresis~\cite{CMS1,CMS2,CMS3,CMS4,CMS5}, specific development and optimization of intricate processes are required for different chiral molecular species.
Based on the cyclic three-level model, there are some theoretical  methods~\cite{PRL.87.183002,PRL.90.033001,PRA.77.015403,PRL.99.130403,JCP.132.194315,PRL.122.173202} available to all chiral molecular species for achieving the enantio-pure samples. Unfortunately, the enantio-pure samples can not be obtained by using these methods~\cite{PRL.87.183002,PRL.90.033001,PRA.77.015403,PRL.99.130403,JCP.132.194315,PRL.122.173202} in the existing experimental technical conditions~\cite{JCP.137.044313,PRL.118.123002}.
Therefore, methods for the determination of enantiomeric excess only based on electric-dipole transitions and  without requiring the enantio-pure samples are highly desired.

In this paper, we theoretically propose such a method for
the determination of enantiomeric excess based on chirality-dependent
AC Stark effects in the cyclic three-level model.
Three electromagnetic fields are applied to couple the three electric-dipole transitions under the three-photon resonance. Around each of the three bare transition frequencies, six peaks with
chirality-dependent frequency shifts will appear in the Fourier transform spectrum of the induced polarization for each of the two enantiomers as the result of AC Stark effects.
The amplitudes of these AC Stark peaks for each enantiomers are (approximately) proportional to the molecule numbers of that enantiomer in the chiral mixture. Based on these properties, we give a new method for determination of enantiomeric excess with
the help of a racemic sample. Since the racemic sample is easy to get, our new method would have wide applications in the determinations of enantiomeric excess for chiral molecules whose enantio-pure samples are still challenging to achieve.

\section{Model}
\begin{figure}[h]
  \centering
  \includegraphics[width=0.7\columnwidth]{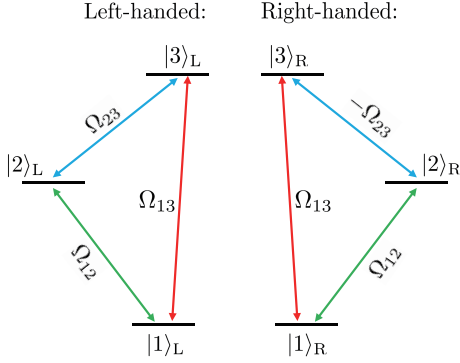}\\
  \caption{Modeling the left- and right-handed chiral molecules as cyclic three-level systems. Three electromagnetic fields couple to the transitions of the cyclic three-level systems with Rabi frequencies $\Omega_{12}$, $\Omega_{13}$, and $\pm\Omega_{23}$.}\label{Fig1}
\end{figure}

Our method is based on the cyclic three-level model~\cite{PRL.87.183002,PRL.90.033001,PRA.77.015403,PRL.99.130403,JCP.132.194315,PRA.98.063401} of chiral molecules formed with three electric-dipole transitions. The general cyclic three-level model is described with the following Hamiltonian in the rotating-wave approximation ($\hbar=1$):
\begin{align}\label{HI}
\hat{\mathrm{H}}=\sum^{3}_{j=1}  v_{j}|j\rangle\langle j|
+\sum^{3}_{i>j=1}[ \Omega_{ji}e^{\mathrm{i}\omega_{ji}t}|j\rangle\langle i|+h.c.],
\end{align}
where $  v_{j}$ are the energies of the states $|j\rangle$, $\omega_{ji}$ are the
frequencies of the electromagnetic fields coupling to the transitions
$|i\rangle\leftrightarrow|j\rangle$, and the Rabi frequencies $\Omega_{ji}$ can be
controlled by varying the corresponding electromagnetic fields. We
will specify our model by choosing the Rabi frequencies of the left- and right- handed molecules
as $\Omega^{\mathrm{L}}_{ji}=\Omega_{ji}$, $\Omega^{\mathrm{R}}_{12}=\Omega_{12}$, $\Omega^{\mathrm{R}}_{13}=\Omega_{13}$,
and $\Omega^{\mathrm{R}}_{23}=-\Omega_{23}$ as shown in Fig.~\ref{Fig1}.
The inner states of the left- and right-handed molecules
are $|j\rangle_{\mathrm{L}}$ and $|j\rangle_{\mathrm{R}}$ ($j=1,2,3$). We have introduced the
scripts $\mathrm{L}$ and $\mathrm{R}$ to denote the left- and right-handed molecules.

Under the three-photon resonance ($\omega_{13}-\omega_{12}=\omega_{23}$), we write
the evolved wave function for a general cyclic three-level system governed by Eq.~(\ref{HI}) as
\begin{align}\label{STT}
|\Psi(t)\rangle&=e^{-\mathrm{i} v_{1}t}[c_{1}(t)|1\rangle+\sum^{3}_{j=2}c_{j}(t)e^{-\mathrm{i}\omega_{1j}t}|j\rangle],
\end{align}
where the column vector $\bm{c}=(c_1,c_2,c_3)^{\mathrm{T}}$ can be evaluated from the following equation
\begin{align}\label{SDEQ}
\mathrm{i}\dot{\bm{c}}(t)=\mathcal{H}\cdot\bm{c}(t)
\end{align}
with the time-independent matrix
\begin{align}\label{MH}
\mathcal{H}=
\left(
  \begin{array}{ccc}
    0 & \Omega_{12} & \Omega_{13} \\
    \Omega^{\ast}_{12} & \Delta_{12} & \Omega_{23} \\
    \Omega^{\ast}_{13} & \Omega^{\ast}_{23} & \Delta_{13} \\
  \end{array}
\right)
\end{align}
in the basis $\{|1\rangle,|2\rangle,|3\rangle\}$.
Here the detunings are defined as $\Delta_{ji}\equiv v_i-v_j-\omega_{ji}$ with $3\ge i>j\ge1$, which
satisfy the three-photon resonance condition $\Delta_{12}+\Delta_{23}=\Delta_{13}$.
The evolved wave function $|\Psi(t)\rangle$ in Eq.~(\ref{STT}) is governed by the evolution
of $\bm{c}(t)$ with
\begin{align}\label{CE}
\bm{c}(t)=\sum^{3}_{j=1}\alpha_{j}e^{-\mathrm{i}\xi_{j}t}\bm{\eta}_{j},
\end{align}
where $\bm{\eta}_{j}=({\eta}_{j1},{\eta}_{j2},{\eta}_{j3})^{\mathrm{T}}$ and $\xi_{j}$ are the eigenvectors and the corresponding eigenvalues of the matrix~(\ref{MH}). The exact forms of $\bm{\eta}_{j}$ and $\xi_{j}$
can be found in Refs.~\cite{PRL.87.183002,SCP}. The coefficients $\alpha_{j}$ are determined by the initial state.

\section{AC Stark effects}\label{CDS}
The AC Stark effects will emerge in the evolution of the cyclic three-level systems.
In order to demonstrate this, we explore the induced polarization, which is
a transient observable~\cite{anie.201307159}.
For a single molecule, the induced polarization is $\bm{\mathrm{\mathcal{P}}}(t)\equiv \langle\Psi(t)|\hat{\bm{\mu}}|\Psi(t)\rangle=\bm{\mathrm{\mathcal{P}}}_{21}(t)
+\bm{\mathrm{\mathcal{P}}}_{31}(t)+\bm{\mathrm{\mathcal{P}}}_{32}(t)$ with $\hat{\bm{\mu}}$ the electric-dipole operator, where $\bm{\mathrm{\mathcal{P}}}_{21}$,
$\bm{\mathrm{\mathcal{P}}}_{31}$, and $\bm{\mathrm{\mathcal{P}}}_{32}$ correspond to
the transition electric-dipole moments $\bm{\mu}_{21}$, $\bm{\mu}_{31}$, and $\bm{\mu}_{32}$. They are respectively
\begin{align}\label{mu}
&\bm{\mathrm{\mathcal{P}}}_{21}(t)=\sum^{3}_{j_1,j_2=1}[\Gamma_{21}(j_{1},j_{2})
e^{\mathrm{i}(\omega_{12}+\xi_{j_{1}}-\xi_{j_{2}})t}\bm{\mu}_{21}+c.c.],\nonumber\\
&\bm{\mathrm{\mathcal{P}}}_{31}(t)=\sum^{3}_{j_1,j_2=1}[\Gamma_{31}(j_{1},j_{2})
e^{\mathrm{i}(\omega_{13}+\xi_{j_{1}}-\xi_{j_{2}})t}\bm{\mu}_{31}+c.c.],\nonumber\\
&\bm{\mathrm{\mathcal{P}}}_{32}(t)=\sum^{3}_{j_1,j_2=1}[\Gamma_{32}(j_{1},j_{2})
e^{\mathrm{i}(\omega_{23}+\xi_{j_{1}}-\xi_{j_{2}})t}\bm{\mu}_{32}+c.c.].
\end{align}
Here the time-independent terms are
\begin{align}\label{GM}
\Gamma_{j^{\prime}_1 j^{\prime}_2}(j_{1},j_{2})=\alpha_{j_{2}}\alpha^{\ast}_{j_{1}}
\eta_{j_{2}j^{\prime}_2}\eta^{\ast}_{j_{1}j^{\prime}_1}
\end{align}
with $3\ge j^{\prime}_1 >j^{\prime}_2\ge1$ and $\bm{\mu}_{ij}\equiv\langle i|\bm{\hat{\mu}}|j\rangle$.

From Eq.~(\ref{mu}), we find that, around each bare transition frequency for the cyclic three-level system, there will be
six peaks in the Fourier transform spectrum of the induced polarization with frequency shifts of
\begin{align}\label{DF}
\delta\omega_{j_1j_2}=(\xi_{j_{1}}-\xi_{j_{2}}),~~~ j_1\ne j_2.
\end{align}
Such frequency shifts are known as AC Stark shifts~\cite{PR.100.703}. In the following, we
will call the peaks due to the AC Stark shifts as AC Stark peaks.

Generally, the frequency shifts~(\ref{DF}) are chirality-dependent due to the chirality-dependency of $\mathcal{H}$~\cite{Note1}.
In the following, we will show the chirality-dependency of the frequency shifts in a specific case of the initial ground state $|1\rangle$, i.e., $\bm{c}(0)=(1,0,0)^{\mathrm{T}}$, and the parameters are tuned to satisfy the conditions
\begin{align}\label{CD1}
\Omega_{13}=\Omega_{12}\ne0,~~
\Delta_{12}=\Delta_{13}.
\end{align}
Under the conditions~(\ref{CD1}), the eigenvectors of the matrix~(\ref{MH}) are
\begin{align}
&\bm{\eta}_{1}=\frac{\sqrt{2}}{2}(0,-1,1)^{\mathrm{T}},\nonumber\\ &\bm{\eta}_{2}=\mathcal{N}_2(-\frac{\xi_{3}}{\Omega_{12}},1,1)^{\mathrm{T}},\nonumber\\ &\bm{\eta}_{3}=\mathcal{N}_3(-\frac{\xi_{2}}{\Omega_{12}},1,1)^{\mathrm{T}}
\end{align}
with the normalization coefficients $\mathcal{N}_{2}$ and $\mathcal{N}_{3}$, and the corresponding
eigenvalues $\xi_{1}=\Delta_{12}-\Omega_{23}$, $\xi_{2}=[\Delta_{12}+\Omega_{23}-\sqrt{8\Omega^2_{12}+(\Delta_{12}+\Omega_{23})^2}]/2$, and  $\xi_{3}=[\Delta_{12}+\Omega_{23}+\sqrt{8\Omega^2_{12}+(\Delta_{12}+\Omega_{23})^2}]/2$.
Here, we have assumed that all the Rabi frequencies are real for simplicity.

Note that $\bm{c}(0)=(1,0,0)^{\mathrm{T}}$ is orthogonal to $\bm{\eta}_1$.
According to Eq.~(\ref{GM}), the six AC Stark peaks will reduce to two AC Stark peaks for
each of the two enantiomers with the frequency shifts
\begin{align}\label{DO}
&\delta\omega^{\mathrm{L}}_{32}=\sqrt{8\Omega^2_{12}+(\Delta_{12}+\Omega_{23})^2},
~~\delta\omega^{\mathrm{L}}_{23}=-\delta\omega^{\mathrm{L}}_{32}\nonumber\\
&\delta\omega^{\mathrm{R}}_{32}=\sqrt{8\Omega^2_{12}+(\Delta_{12}-\Omega_{23})^2},
~~\delta\omega^{\mathrm{R}}_{23}=-\delta\omega^{\mathrm{R}}_{32}.
\end{align}
Here, the
scripts $\mathrm{L}$ and $\mathrm{R}$ stand for the left-and right-handed molecules.
The distance between the
two adjacent AC Stark peaks corresponding to the two enantiomers is
\begin{align}\label{DST}
d\equiv\left|\delta\omega^{\mathrm{L}}_{32}-\delta\omega^{\mathrm{R}}_{32}\right|
=\left|\delta\omega^{\mathrm{L}}_{23}-\delta\omega^{\mathrm{R}}_{23}\right|.
\end{align}
From Eq.~(\ref{DO}) and Eq.~(\ref{DST}), we find that, in the parameter
regime $|\Omega_{12}|\gg\{|\Delta_{12}|,|\Omega_{23}|\}$, $d$ will approach $zero$.
In what follows we will focus on the parameter regime
$|\Omega_{12}|\le\{|\Delta_{12}|,|\Omega_{23}|\}$, where
$d$ can be sufficient large and thus the AC Stark peaks corresponding to the two enantiomers
clearly distinguishable.

It is worthy to note that the chirality-dependency of the AC Stark shifts results from the
fact that the product of the three electric-dipole transition moments
for the three transitions changes sign for different enantiomers, which is also
the essential in the EMS method~\cite{Hirota,Nature.497.475,PRL.111.023008,PCCP.16.11114,
ACI,JCP.142.214201,JPCL.6.196,JPCL.7.341,Angew.Chem.10.1002,KK}. In the EMS method~\cite{Hirota,Nature.497.475,PRL.111.023008,PCCP.16.11114,
ACI,JCP.142.214201,JPCL.6.196,JPCL.7.341,Angew.Chem.10.1002,KK}, only two transitions are
driven by electromagnetic fields. In our scheme, the three transitions
are driven by three corresponding electromagnetic fields. Such a scheme with three electromagnetic fields
has also been used for enantio-separation~\cite{PRL.87.183002,PRL.90.033001,PRA.77.015403,PRL.118.123002}.

\section{Method for determination of enantiomeric excess}

Now, we have demonstrated the chirality-dependent AC Stark peaks around each of the three bare transition frequencies in the Fourier transform spectrum of the induced polarization. Such chirality-dependent properties can be used for the determination of enantiomeric excess.

\subsection{Ideal cases}
In the ideal cases without decoherence,
the induced polarization corresponding to $\bm{\mu}_{21}$ for a chiral mixture with $N_{\mathrm{L}}$ left-handed molecules and $N_{\mathrm{R}}$ right-handed ones is given in the frequency domain as~\cite{Hirota,Nature.497.475,PRL.111.023008,PCCP.16.11114,
ACI,JCP.142.214201,JPCL.6.196,JPCL.7.341,Angew.Chem.10.1002,KK}
\begin{align}\label{P21}
\mathbf{P}_{21}(\omega)=
N_{\mathrm{L}}\bm{\mathrm{\mathcal{P}}}^{\mathrm{L}}_{21}(\omega)+N_{\mathrm{R}}\bm{\mathrm{\mathcal{P}}}^{\mathrm{R}}_{21}(\omega),
\end{align}
where $\mathbf{P}_{21}(\omega)$, $\bm{\mathrm{\mathcal{P}}}^{\mathrm{L}}_{21}(\omega)$, and $\bm{\mathrm{\mathcal{P}}}^{\mathrm{R}}_{21}(\omega)$ are Fourier transforms of $\mathbf{P}_{21}(t)$, $\bm{\mathrm{\mathcal{P}}}^{\mathrm{L}}_{21}(t)$, and $\bm{\mathrm{\mathcal{P}}}^{\mathrm{R}}_{21}(t)$.
Here, the coherent superposition of the induced polarizations in Eq.~(\ref{P21}) is
essentially important for our discussions, which is also the starting point of current experiments using the EMS method~\cite{Hirota,Nature.497.475,PRL.111.023008,PCCP.16.11114,
ACI,JCP.142.214201,JPCL.6.196,JPCL.7.341,Angew.Chem.10.1002,KK}. The succusses of the related experiments~~\cite{Nature.497.475,PRL.111.023008,PCCP.16.11114,
ACI,JCP.142.214201,JPCL.6.196,JPCL.7.341,Angew.Chem.10.1002}
ensure the validity of Eq.~(\ref{P21}) in current experiments conditions~~\cite{Nature.497.475,PRL.111.023008,PCCP.16.11114,
ACI,JCP.142.214201,JPCL.6.196,JPCL.7.341,Angew.Chem.10.1002}.
One approximation applied to reduce the complexity is that the interaction volume of the fields and molecules is small compared to the wavelength by neglecting the dependence of the electromagnetic field on the wave vectors~\cite{Angew.Chem.10.1002}.

The key point of our method is to find two frequencies $\omega_{\mathrm{I}}$ and
$\omega_{\mathrm{II}}$, where the induced polarizations are proportional to
the molecule numbers of the two enantiomers respectively:
\begin{align}\label{LE}
&\mathbf{P}_{21}(\omega_{\mathrm{I}})=N_{\mathrm{L}}\bm{\mathrm{\mathcal{P}}}^{\mathrm{L}}_{21}(\omega_{\mathrm{I}}),\nonumber\\
&\mathbf{P}_{21}(\omega_{\mathrm{II}})=N_{\mathrm{R}}\bm{\mathrm{\mathcal{P}}}^{\mathrm{R}}_{21}(\omega_{\mathrm{II}}).
\end{align}
In the ideal cases, the AC Stark peaks are
width-free according to Eq.~(\ref{mu}).
Thus, we can choose the two frequencies corresponding to two peaks of
the two enantiomers respectively with $\bm{\mathrm{\mathcal{P}}}^{\mathrm{R}}_{21}(\omega_{\mathrm{I}})=0$ and $\bm{\mathrm{\mathcal{P}}}^{\mathrm{L}}_{21}(\omega_{\mathrm{II}})=0$.

For the enantiomeric excess $\varepsilon\equiv(N_{\mathrm{L}}-N_{\mathrm{R}})/(N_{\mathrm{L}}+N_{\mathrm{R}})$, which defines the excess of one enantiomer over the other in the mixture, we provide a new method to determine it through
\begin{align}\label{MEE}
\varepsilon=\frac{|\mathbf{P}_{21}(\omega_{\mathrm{I}})|-\lambda |\mathbf{P}_{21}(\omega_{\mathrm{II}})|}{|\mathbf{P}_{21}(\omega_{\mathrm{I}})|+\lambda |\mathbf{P}_{21}(\omega_{\mathrm{II}})|}
\end{align}
with
\begin{align}\label{LB}
\lambda=\frac{|\bm{\mathrm{\mathcal{P}}}^{\mathrm{L}}_{21}(\omega_{\mathrm{I}})|}
{|\bm{\mathrm{\mathcal{P}}}^{\mathrm{R}}_{21}(\omega_{\mathrm{II}})|}.
\end{align}
Here the coefficient $\lambda$ can be calculated with the help of Eq.~(\ref{GM}) for the left- and right-handed molecules in the ideal cases.

\subsection{Realistic cases}\label{RCD}
In the realistic cases where the effect of decoherence should be taken into consideration,
the peaks on the Fourier spectrum will be broadened and the linear relationships between the
induced polarizations and the molecule numbers in Eq.~(\ref{LE})
will change as~\cite{Nature.497.475,PRL.111.023008,PCCP.16.11114,ACI,
JCP.142.214201,JPCL.6.196,JPCL.7.341,Angew.Chem.10.1002,anie.201307159}
\begin{align}\label{P1}
&\mathbf{\tilde{P}}_{21}(\omega_{\mathrm{I}})=N_{\mathrm{L}}\bm{\mathrm{\mathcal{\tilde{P}}}}^{\mathrm{L}}_{21}(\omega_{\mathrm{I}})
+N_{\mathrm{R}}\bm{\mathrm{\mathcal{\tilde{P}}}}^{\mathrm{R}}_{21}(\omega_{\mathrm{I}}),\nonumber\\
&\mathbf{\tilde{P}}_{21}(\omega_{\mathrm{II}})=N_{\mathrm{L}}\bm{\mathrm{\mathcal{\tilde{P}}}}^{\mathrm{L}}_{21}(\omega_{\mathrm{II}})
+N_{\mathrm{R}}\bm{\mathrm{\mathcal{\tilde{P}}}}^{\mathrm{R}}_{21}(\omega_{\mathrm{II}}),
\end{align}
where $\mathbf{\tilde{P}}_{21}$, $\bm{\mathrm{\mathcal{\tilde{P}}}}^{\mathrm{L}}_{21}$, and $\bm{\mathrm{\mathcal{\tilde{P}}}}^{\mathrm{R}}_{21}$ 
stand for the induced polarizations in the realistic cases.

In order to determine the enantiomeric excess,
we should adjust the parameters and appropriately choose the two AC Stark
peaks for determination of enantiomeric excess
to ensure that, at the frequencies $\omega_{\mathrm{I}}$ and
$\omega_{\mathrm{II}}$ of the two chosen AC Stark peaks,
the induced polarizations are (approximately) proportional to
the molecule numbers of the two enantiomers respectively.
Specifically, the parameters should be adjusted to ensure that $|\bm{\mathrm{\mathcal{\tilde{P}}}}^{\mathrm{\mathrm{R}}}_{21}
(\omega_{\mathrm{I}})|\ll|\bm{\mathrm{\mathcal{\tilde{P}}}}^{\mathrm{L}}_{21}(\omega_{\mathrm{I}})$
and $|\bm{\mathrm{\mathcal{\tilde{P}}}}^{\mathrm{L}}_{21}(\omega_{\mathrm{II}})|\ll
|\bm{\mathrm{\mathcal{\tilde{P}}}}^{\mathrm{R}}_{21}(\omega_{\mathrm{II}})|$.
Generally, such properties can be realized when the two chosen AC Stark peaks
are well separated comparing with the widths of the peaks due to the decoherence.
Thus, we can estimate the
enantiomeric excess as
\begin{align}\label{MEE1}
\varepsilon\simeq\varepsilon_{m}\equiv\frac{|\mathbf{\tilde{P}}_{21}(\omega_{\mathrm{I}})|-\lambda |\mathbf{\tilde{P}}_{21}(\omega_{\mathrm{II}})|}{|\mathbf{\tilde{P}}_{21}(\omega_{\mathrm{I}})|+\lambda |\mathbf{\tilde{P}}_{21}(\omega_{\mathrm{II}})|}.
\end{align}
In the realistic case, the coefficient $\lambda$ can be measured with the help of the Fourier transform spectrum of a racemic mixture as
\begin{align}\label{lb1}
\lambda\simeq\lambda_{m}\equiv\frac{|\mathbf{\tilde{P}}^{\mathrm{rm}}_{21}(\omega_{\mathrm{I}})|}{|\mathbf{\tilde{P}}^{\mathrm{rm}}_{21}(\omega_{\mathrm{II}})|},
\end{align}
where ``$rm$'' denotes that the sample is a racemic mixture.

\subsection{Heterodyne measurement}
The induced polarization $\mathbf{\tilde{P}}_{21}$ (or $\mathbf{{P}}_{21}$) can be experimentally determined by detecting the induced electromagnetic field $\bm{E}_{21}(t)$ at point $\bm{r}$ in the radiation zone~\cite{Book2,1809.02646}. The component of $\bm{E}_{21}(t)$ with the frequency $\omega$ is
\begin{align}
\bm{E}_{21}(\omega)= Z_{0}\frac{c k^2}{4\pi}\frac{e^{\mathrm{i}kr}}{r^3}(\bm{r}\times\mathbf{\tilde{P}}_{21}(\omega))\times\bm{r},
\end{align}
where $Z_{0}=\sqrt{\mu_0/\varepsilon_0}$, $r=|\bm{r}|$, and $k$ is the wave-number corresponding to the frequency $\omega$. Usually, the signal $\bm{E}_{21}$ is weak. We may amplify it
in the case of heterodyne measurement with a strong local oscillator $\bm{E}_{\mathrm{lo}}$ with the frequency $\omega_{\mathrm{lo}}$.
This is realized by measuring the cross term between a weak signal and strong local oscillator as
\begin{align}\label{HM}
\delta I(t)&=\left|\bm{E}_{21}(t)+\bm{E}_{\mathrm{lo}}(t)\right|^2
-\left|\bm{E}_{\mathrm{lo}}(t)\right|^2\nonumber\\
&\simeq 2 \left|\bm{E}_{21}(t)\cdot\bm{E}_{\mathrm{lo}}(t)\right|
\end{align}
with $\left|\bm{E}_{21}\right|\ll\left|\bm{E}_{\mathrm{lo}}\right|$.
For convenience, we have chosen the polarization direction of $\bm{E}_{\mathrm{lo}}$ along that of $\bm{E}_{21}$.
The Fourier transform of $\delta I(t)$ will be in essence the same as the Fourier transform of
the induced polarization $\mathbf{\tilde{P}}_{21}$.
In practice, we can use a grating to spatially disperse the frequency components of the signal, which is in essence similar to Fourier transform.
We note that $\mathbf{\tilde{P}}_{21}$, $\mathbf{\tilde{P}}_{31}$,
and $\mathbf{\tilde{P}}_{32}$ are mutually orthogonal to
each other. The corresponding induced electromagnetic fields due to
$\mathbf{\tilde{P}}_{31}$ and $\mathbf{\tilde{P}}_{32}$
will be orthogonal to $\bm{E}_{\mathrm{lo}}$, and thus will not affect the heterodyne measurement
of $\bm{E}_{21}$ according to Eq.~(\ref{HM}).

\section{Examples of 1,2-propanediol}
Specifically, we will use $1,2$-propanediol as an example to demonstrate our method.
The working states of the cyclic three-level model are three rotational states of the ground vibrational state. They can be $|1\rangle=|0_{0,0}\rangle$, $|2\rangle=|1_{-1,0}\rangle$, and $|3\rangle=(|1_{1,1}\rangle+|1_{1,-1}\rangle)/\sqrt{2}$ in the $|J_{\tau,M}\rangle$ notation with the angular moment quantum number $J$, the magnetic quantum number $M$, and $\tau$ running from $-J$ to
$J$ in unit steps in the order of increasing energy~\cite{PRA.98.063401}. The bare transition frequencies are
$\omega_{12}=6.4$\,GHz, $\omega_{13}=12.2$\,GHz, and $\omega_{23}=5.8$\,GHz.

Three linearly $Z$-, $Y$-, and $X$- polarized electromagnetic fields are applied to couple the three transitions $|1\rangle\leftrightarrow|2\rangle$, $|1\rangle\leftrightarrow|3\rangle$, and $|2\rangle\leftrightarrow|3\rangle$, respectively. We assume the
initial state of each of the two enantiomers is the ground state $|1\rangle$.
The system can be described with the cyclic three-level model~\cite{PRA.98.063401}.

The electric dipoles in the molecular frame for the three transitions $|1\rangle\leftrightarrow|2\rangle$,
$|1\rangle\leftrightarrow|3\rangle$, and $|2\rangle\leftrightarrow|3\rangle$
are $\mu_{z}=1.201$\,D, $\mu_{y}=0.365$\,D, and $\mu_{x}=1.916$\,D, respectively~\cite{PRL.118.123002}.
In the space-fixed frame, we have $|\bm{\mu}_{12}|=\mu_{z}/\sqrt{3}=0.69$\,D, $|\bm{\mu}_{13}|=\mu_{y}/\sqrt{3}=0.21$\,D,
and $|\bm{\mu}_{23}|=\mu_{x}/2=0.96$\,D~\cite{PRA.98.063401}.
The intensities of the electromagnetic fields are limited in the current
experimental condition~\cite{PRL.118.123002,Nature.497.475,PRL.111.023008}.
In our simulations, we will choose the maximum value of the intensities of
the electromagnetic fields to be $\sim10$\,V/cm.
Then the corresponding Rabi frequencies can be tuned in the regimes
$|\Omega_{12}|\lesssim22$\,MHz, $|\Omega_{13}|\lesssim10$\,MHz,
and $|\Omega_{23}|\lesssim30$\,MHz.


\subsection{Effect of decoherence: Master equation}

In appearance of decoherence, we should use the density matrix $\hat{\rho}=\sum^{3}_{l,u=1}\rho_{lu}(t)e^{\mathrm{i}\omega_{lu}t}|l\rangle\langle u|$ to depict the state of a single
general three-level system and thus we have
\begin{align}\label{PTC}
\bm{\mathcal{\tilde{P}}}_{21}(t)=\rho_{12}(t)e^{\mathrm{i}\omega_{12}t}\bm{\mu}_{21}+c.c..
\end{align}
The evolution of the density matrix is described by the master equations~\cite{JPB.37.2811}
\begin{align}\label{MST}
\frac{d {\rho}}{d t}=-\mathrm{i}[\mathcal{{H}},{\rho}]+\mathcal{{L}}({\rho}).
\end{align}
Explicitly, $\mathcal{{L}}({\rho})$ representing the decoherence can be written as~\cite{JPB.37.2811}
\begin{align}
&[\mathcal{{L}}({\rho})]_{lu}
=-(\gamma_{lu}+\gamma^{\mathrm{ph}}_{lu})\rho_{lu},~~~~(l\ne u)\nonumber\\
&[\mathcal{{L}}({\rho})]_{uu}=
\sum^{3}_{u^{\prime}=u+1}\Gamma_{uu^{\prime}}\rho_{u^{\prime}u^{\prime}}
-\sum^{u-1}_{u^{\prime}=1}\Gamma_{u^{\prime}u}\rho_{uu},
\end{align}
where
\begin{align}
\gamma_{lu}=\frac{\gamma_{l}+\gamma_{u}}{2},
~~~~\gamma_{u}=\sum^{u-1}_{l^{\prime}=1}\Gamma_{l^{\prime}u}.
\end{align}
$\Gamma_{lu}$ ($l\ne u$) are taken into account for pure population relaxation due to radiative
processes and the effect of collisional processes on population relaxation from one level to
another. The effect of collisions on pure dephasing is depicted by $\gamma^{\mathrm{ph}}_{lu}$.

For simplicity, we assume that
\begin{align}
\Gamma_{lu}=\Gamma,~~~~\gamma^{\mathrm{ph}}_{lu}=\gamma,
\end{align}
which give $\gamma_{12}={\Gamma}/{2},~~\gamma_{13}=\Gamma,~~\gamma_{23}={3}\Gamma/2$.
For chiral molecules in the gaseous medium, the decoherence is dominated
by inelastic collisions and thus the pure dephasing due to the elastic collisions is
negligible~\cite{JCP.126.034503,JCP.129.134301}.
In Ref.~\cite{MP.110.1757}, it is also mentioned that $\Gamma>\gamma$
in buffer gas medium. Further, we assume $\gamma=0$ and
$\Gamma\simeq0.1$\,MHz~\cite{PRL.118.123002,MP.110.1757} in the following calculations.

\subsection{Fourier transform spectrum}

\begin{figure}[htp]
  \centering
  \includegraphics[width=0.9\columnwidth]{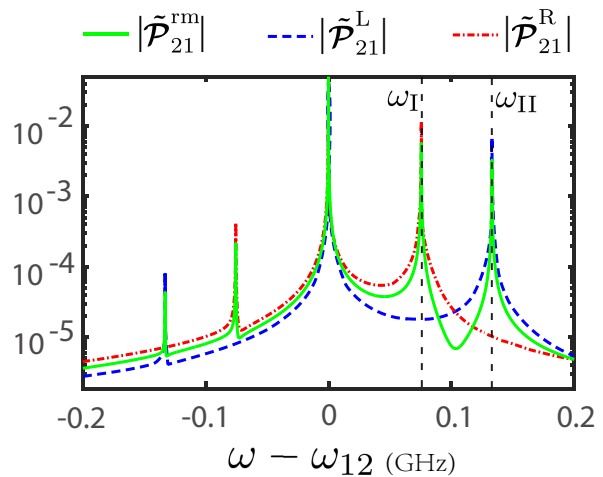}\\
  \caption{Fourier transform spectrum of average induced polarization in a racemic sample $|\bm{\mathrm{\mathcal{\tilde{P}}}}^{\mathrm{rm}}_{21}(\omega)|$ (green line), single left-handed molecule $|\bm{\mathrm{\mathcal{\tilde{P}}}}^{\mathrm{L}}_{21}(\omega)|$ (blue line), and single right-handed molecule $|\bm{\mathrm{\mathcal{\tilde{P}}}}^{\mathrm{R}}_{21}(\omega)|$ (red line) in the initial ground state. The two AC Stark peaks corresponding to frequencies
  $\omega_{\mathrm{I}}$ and $\omega_{\mathrm{II}}$ are chosen for the determination of enantiomeric excess.
  The parameters are $\Omega_{12}=\Omega_{13}=10$\,MHz,
  $\Omega_{23}=30$\,MHz, and $\Delta_{12}=\Delta_{13}=100$\,MHz. According to the
  three-photon resonance condition, we have $\Delta_{23}=0$\,MHz. The time period
  for calculation of fast Fourier transform is $0<t<1$\,ms.}\label{Fig2}
\end{figure}

For simplicity, we will assume that all parameters are positive and focus in the regime $\Delta_{12}>\Omega_{23}>0$.
In Fig.~\ref{Fig2}, we show the Fourier transform spectrum of \begin{align}
\bm{\mathrm{\mathcal{\tilde{P}}}}^{\mathrm{rm}}_{21}(t)\equiv\frac{\mathbf{\tilde{P}}^{\mathrm{rm}}_{21}(t)}{N}
\end{align}
for the average induced polarization corresponding to $\bm{\mu}_{21}$ of a racemic mixture with $N$ molecules in the initial ground state (green line). It is worthy to note that the molecule number $N$ is not required to be equal to the molecule number of the chiral mixture to be detected $(N_{L}+N_{R})$.
The calculation period is  $0<t<1$\,ms.
The parameters are
$\Omega_{12}=\Omega_{13}=10$\,MHz,
$\Omega_{23}=30$\,MHz, $\Delta_{12}=\Delta_{13}=100$\,MHz, and $\Delta_{23}=0$\,MHz
under the three-photon resonance.
The chirality-dependency in Eq.~(\ref{mu}) is related to
the evolutions of the three-level systems. In order to have obvious evolutions, we have chosen the maximum values of $\Omega_{13}$ and $\Omega_{23}$.
Here, we note that the working states are appropriately
chosen to make sure that the parameters can
be tuned in the parameter regime $|\Omega_{12}|\le\{|\Delta_{12}|,|\Omega_{23}|\}$,
where $d$ can be such large that our method~(\ref{MEE})
can give a good estimation of the enantiomeric excess in the realistic cases.
\begin{figure*}[htp]
  \centering
  \includegraphics[width=1.9\columnwidth]{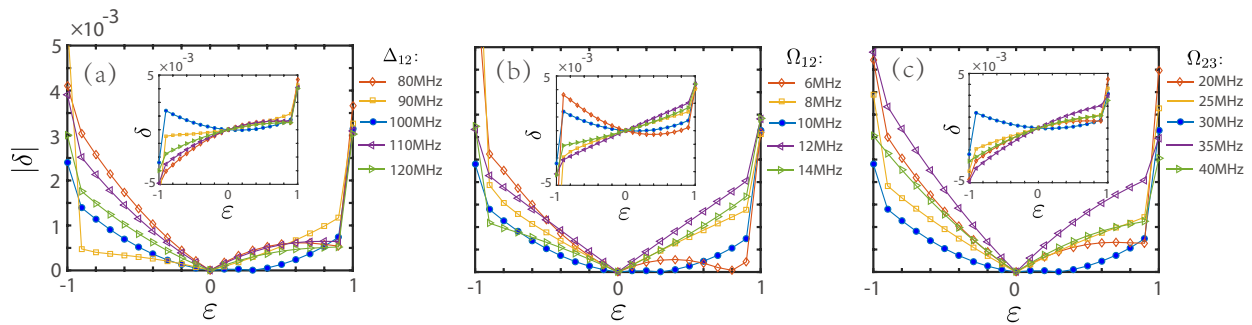}\\
  \caption{Absolute values of the errors $\delta=\varepsilon-\varepsilon_{m}$ as a function of $\varepsilon$: (a) for different $\Delta_{12}$ with $\Omega_{12}=\Omega_{13}=10$\,MHz,
  $\Omega_{23}=30$\,MHz, and $\Delta_{13}=\Delta_{12}$; (b) for different $\Omega_{12}$ with
  $\Omega_{13}=\Omega_{12}$, $\Omega_{23}=30$\,MHz, and
  $\Delta_{12}=\Delta_{13}=100$\,MHz; (c) for different $\Omega_{23}$ with
  $\Omega_{12}=\Omega_{13}=10$\,MHz and $\Delta_{13}=\Delta_{12}=100$\,MHz. For the three cases, we have $\Delta_{23}=0$\,MHz under the
  three-photon resonance condition.  }\label{Fig3}
\end{figure*}
The average induced polarization corresponding $\bm{\mathrm{\mathcal{\tilde{P}}}}^{\mathrm{rm}}_{21}(t)=0.5\bm{\mathrm{\mathcal{\tilde{P}}}}^{\mathrm{L}}_{21}(t)
+0.5\bm{\mathrm{\mathcal{\tilde{P}}}}^{\mathrm{R}}_{21}(t)$ can be obtained by numerically solving Eq.~(\ref{PTC}) and Eq.~(\ref{MST}). After that, using a fast Fourier transform algorithm, we arrive the Fourier transform spectrum of $\bm{\mathrm{\mathcal{\tilde{P}}}}^{\mathrm{rm}}_{21}(t)$. We find that
the six AC Stark peaks for each
of the two enantiomers reduce to two AC Stark peaks.
One of the two AC Stark peaks for each of the two enantiomers
is suppressed. The unsuppressed AC Stark peaks are at the same side of the bare transition frequency according to Eq.~(\ref{GM}). It is known from Eq.~(\ref{DO}) that, the two unsuppressed AC Stark peaks correspond to the two enantiomers, respectively. Their corresponding frequencies are $\omega_{\mathrm{I}}$ and $\omega_{\mathrm{II}}$. We also give the Fourier transform spectrums of
$\bm{\mathrm{\mathcal{\tilde{P}}}}^{\mathrm{L}}_{21}$ (blue line) and $\bm{\mathrm{\mathcal{\tilde{P}}}}^{\mathrm{R}}_{21}$ (red line) in Fig.~\ref{Fig2}, which clearly show $|\bm{\mathrm{\mathcal{\tilde{P}}}}^{\mathrm{R}}_{21}
(\omega_{\mathrm{I}})|\ll|\bm{\mathrm{\mathcal{\tilde{P}}}}^{\mathrm{L}}_{21}(\omega_{\mathrm{I}})|$
and $|\bm{\mathrm{\mathcal{\tilde{P}}}}^{\mathrm{L}}_{21}(\omega_{\mathrm{II}})|\ll
|\bm{\mathrm{\mathcal{\tilde{P}}}}^{\mathrm{R}}_{21}(\omega_{\mathrm{II}})|$. Then, the two unsuppressed AC Stark peaks can be considered well separated from other peaks comparing with their widths and Eq.~(\ref{MEE1}) will give a good estimation of the enantiomeric excess by choosing the two AC Stark peaks with frequencies $\omega_{\mathrm{I}}$~and~$\omega_{\mathrm{II}}$.

\subsection{Determination of enantiomeric excess}

In practice,
we can first use a racemic sample and adjust the parameters to ensure that
the two chosen AC Stark peaks are well separated from other peaks and estimate the coefficient $\lambda$ according to Eq.~(\ref{lb1}).
Then applying the same electromagnetic fields to the chiral mixture to be detected,
we can determine the enantiomeric excess from the Fourier transform spectrum of the induced polarization with the help of Eq.~(\ref{MEE1}). Comparing with the EMS method where the enantio-pure sample should have the same molecule number as the chiral mixture to be detected, there is no such a requirement for the racemic sample in our method.

In Fig.~\ref{Fig3}, we numerically show the error
\begin{align}
\delta=\varepsilon-\varepsilon_{m}
\end{align}
as a function of $\varepsilon$ for different parameters.
Our numerical results show that the method~(\ref{MEE1}) gives
a better results in the region $|\varepsilon|\sim0$ than those in the
region $|\varepsilon|\sim1$.
Considering the performance in the entire interval of $\varepsilon$,
we would like to choose the parameters to be
$\Omega_{12}=\Omega_{13}=10$\,MHz, $\Omega_{23}=30$\,MHz, and $\Delta_{12}=\Delta_{13}=100$\,MHz for the estimation of the enantiomeric
excess among the others in Fig.~\ref{Fig3} with the corresponding error of the enantiomeric excess $|\delta|<3\times10^{-3}$.

With our numerical results of initial ground state,
we have demonstrated our method for the determination
of enantiomeric excess and shown that the absolute
error can be deduced by tuning the driving fields.
Generally, in order to give good estimation of the enantiomeric excess,
each of the two chosen AC Stark peaks should be
well separated from the other peaks in the Fourier
transform spectrum, and the two chosen AC Stark peaks
should be sufficiently high comparing with the unchosen peaks.

\section{SUMMARY AND DISCUSSION}
In conclusion, we propose a theoretical method for determination of enantiomeric excess
based on the chirality-dependent AC Stark effects with the framework of the cyclic three-level model
under the three-photon resonance condition. The key point of our method is to find two AC Stark peaks in the Fourier transform spectrum of the induced polarization,
whose amplitudes are (approximately) proportional to the molecule numbers of the two enantiomers, respectively.
Comparing with the traditional methods~\cite{CA1,CA2,CA3} and the EMS method~\cite{Nature.497.475,PRL.111.023008,
PCCP.16.11114,ACI,JCP.142.214201,JPCL.6.196,JPCL.7.341,Angew.Chem.10.1002,KK}, our method determines enantiomeric excess without
requiring for the enantio-pure samples and thus would have wide applications for chiral molecules whose enantio-pure samples are hardly to achieve. The chiral rotational spectroscopy method~\cite{PRA.94.032505}  also work without requiring for the enantio-pure samples. However, the underlying physical mechanism~\cite{PRA.94.032505} is the interference between the
electric- and (usually weak) magnetic-dipole transition moments, which would provide weak chiral signals~\cite{Hirota,Nature.497.475,PRL.111.023008,PCCP.16.11114,ACI,
JCP.142.214201,JPCL.6.196,JPCL.7.341,Angew.Chem.10.1002,KK}.

The relevant physics in our method is the chirality-dependent AC Stark effects. In Ref.~\cite{1501.05282}, they have obtained the chirality-dependent AC Stark effects when all the three transitions are coupled in the large detuning condition. However, the difference of
the frequency shifts for the two enantiomers is very small ($\ll|\Omega_{23}|$). In our proposal,
the difference of the frequency shifts for the two enantiomers is of the same order of $\Omega_{23}$. With the help the chirality-dependent AC Stark effects, one can also determine the enantiomeric excess via power spectrum~\cite{PR.188.1969}, absorption spectrum~\cite{PRA.84.053849}, fluctuation spectrum~\cite{PRA.88.053827}, and so on.\\

\section{Acknowledgement}
This work was supported by the National Key R\&D Program of China grant
(2016YFA0301200), the Natural Science
Foundation of China (under Grants No.~11774024, No.~11534002, No.~U1530401, and No.~U1730449),
and the Science Challenge Project (under Grant No.~TZ2018003).

{}

\end{document}